\documentclass[10pt]{article}

\usepackage[numbers]{natbib}
\usepackage[linkcolor=blue, hidelinks]{hyperref}

\usepackage{graphicx} 
\usepackage{graphics}
\usepackage{color}
\usepackage{url}
\usepackage{subfigure}
\usepackage{latexsym}
\usepackage{amsfonts,amssymb,amsmath}
\usepackage{verbatim}   
\usepackage{amsmath}
\usepackage{amssymb}
\usepackage{fancyhdr}

\usepackage{setspace} 
\usepackage{cleveref}
\usepackage{wrapfig}
\usepackage{etoolbox} 
\usepackage{enumitem}

\usepackage{mdframed}

\crefname{subsection}{section}{sections} 
\crefname{section}{section}{sections} 
\crefname{subsubsection}{section}{sections} 
\usepackage{mathtools}

\usepackage{makecell}
\usepackage{algorithm}
\usepackage{algpseudocode}

\usepackage{pifont}
\newcommand{\xmark}{\ding{55}}
\usepackage{booktabs}
\usepackage{siunitx} 

\newcommand{\R}{\mathbb{R}}

\newcommand{\inv}{^{-1}}

\newcommand{\cv}{\mathbf{c}}

\newcommand{\xv}{\mathbf{x}}
\newcommand{\yv}{\mathbf{y}}

\newcommand{\uvt}[1]{\mathbf{u}(#1)}

\newcommand{\xvt}[1]{\mathbf{x}(#1)}
\newcommand{\yvt}[1]{\mathbf{y}(#1)}

\newcommand{\Av}{\mathbf{A}}

\newcommand{\Cv}{\mathbf{C}}
\newcommand{\Dv}{\mathbf{D}}

\newcommand{\Fv}{\mathbf{F}}
\newcommand{\Gv}{\mathbf{G}}

\newcommand{\Iv}{\mathbf{I}}

\newcommand{\Uv}{\mathbf{U}}
\newcommand{\Vv}{\mathbf{V}}
\newcommand{\Wv}{\mathbf{W}}
\newcommand{\Xv}{\mathbf{X}}
\newcommand{\Yv}{\mathbf{Y}}

\newcommand{\Avt}[1]{\mathbf{A}(#1)}

\newcommand{\Cvt}[1]{\mathbf{C}(#1)}

\newcommand{\Cs}{\mathcal{C}}
\newcommand{\Ds}{\mathcal{D}}

\newcommand{\Ls}{\mathcal{L}}
\newcommand{\Ms}{\mathcal{M}}

\newcommand{\Os}{\mathcal{O}}

\newcommand{\G}{\Gv}
\newcommand{\alphav}{\alpha}
\newcommand{\Wm}{\Wv}
\usepackage{todonotes}
\usepackage{tikz}

\usepackage{xcolor}
\usepackage{amssymb}
\usepackage{scalerel}
\usepackage{stackengine}

\usepackage{multicol}
\usepackage{cleveref}
\usepackage{caption}
\usepackage{authblk}
\usepackage{tabularx} 

\usepackage{tikz,pgfplots}
\usetikzlibrary{positioning}
\pgfdeclarelayer{background}
\pgfsetlayers{background,main}

\tikzstyle{vertex} = [fill,shape=circle,node distance=30pt]
\tikzstyle{edge} = [fill,opacity=.6,fill opacity=.5,line cap=round, line join=round, line width=10pt]
\tikzstyle{elabel} =  [fill,shape=circle,node distance=30pt,fill opacity=.9]

\title{\bf Dynamic Sensor Selection for Biomarker Discovery}

\author[1,*]{Joshua Pickard}
\author[1]{Cooper Stansbury}
\author[2]{Amit Surana}
\author[1]{Lindsey Muir}
\author[3]{Anthony Bloch}
\author[1,3,*]{Indika Rajapakse}
\affil[1]{\small Department of Computational Medicine \& Bioinformatics, University of Michigan, Ann Arbor, MI 48109}
\affil[2]{\small RTX Technology Research Center, East Hartford, CT 06108}
\affil[3]{\small Department of Mathematics, University of Michigan, Ann Arbor, MI 48109}
\affil[*]{\url{{jpc,indikar}@umich.edu}}
\date{} 

\usepackage[a4paper, left=0.5in, right=0.5in, top=0.75in, bottom=0.75in]{geometry}

\begin{document}
\maketitle

\begin{multicols}{2}
\paragraph{Abstract.}

Advances in methods of biological data collection are driving the rapid growth of comprehensive datasets across clinical and research settings.
These datasets provide the opportunity to monitor biological systems in greater depth and at finer time steps than was achievable in the past.
Classically, biomarkers are used to represent and track key aspects of a biological system.
Biomarkers retain utility even with the availability of large datasets, since monitoring and interpreting changes in a vast number of molecules remains impractical.
However, given the large number of molecules in these datasets, a major challenge is identifying the best biomarkers for a particular setting.
Here, we apply principles of observability theory to establish a general methodology for biomarker selection.
We demonstrate that observability measures effectively identify biologically meaningful sensors in a range of time series transcriptomics data.
Motivated by the practical considerations of biological systems, we introduce the method of dynamic sensor selection (DSS) to maximize observability over time, thus enabling observability over regimes where system dynamics themselves are subject to change.
This observability framework is flexible, capable of modeling gene expression dynamics and using auxiliary data, including chromosome conformation, to select biomarkers.
Additionally, we demonstrate the applicability of this approach beyond genomics by evaluating the observability of neural activity.
These applications demonstrate the utility of observability-guided biomarker selection for across a wide range of biological systems, from agriculture and biomanufacturing to neural applications and beyond.

\vspace{2mm}
\noindent
\textbf{Key words}:
\textit{observability} | \textit{biomarkers} | \textit{dynamic sensor selection} | \textit{sensor selection} | \textit{data driven observability} 
\vspace{-2mm}

\section{Introduction}
\noindent
Monitoring the state of a cell or tissue is experimentally and computationally challenging.
Recently developed on-demand sequencing technologies, including live single-cell sequencing and adaptive sampling, are increasing the accessibility of high-dimensional, high-frequency time series genomics data \cite{chen2022live, weilguny2023dynamic}.
These technologies are shifting the bottleneck in monitoring biological systems from the acquisition to the synthesis of data - posing a challenge for the selection of biomarkers that represent a specific biological state in clinical and research settings \cite{califf2018biomarker, ray2010statistical, hartwell2006cancer}.
Observability theory - an engineering framework for sensor selection - provides a framework to uncover biomarkers and offers an approach to analyze and interpret vast biological datasets.

Systems theory models the genome as a dynamical system, where temporal changes of gene expression and chromatin structure are described by the differential equation:
\begin{equation}\label{eq: dynamics}
    \frac{d\xvt{t}}{dt}=f(\xvt{t},\uvt{t}, \theta_f, t).
\end{equation}
The cells state is described by a vector $\xvt{t}\in\R^n,$ environmental influences and perturbations are represented by a control signal $\uvt{t},$ and the function $f(\cdot)$ models the dynamics with parameters $\theta_f.$
Observability involves an additional measurement operator $g(\cdot)$, which maps the system state $\xvt{t}$ to available data $\yvt{t}\in\R^p$ with the equation:
\begin{equation}\label{eq: observer}
    \yvt{t}=g(\xvt{t}, \uvt{t}, \theta_g, t).
\end{equation}
Here, $p$ is the number of measurements collected at each time point, which is often significantly smaller than the dimension $n$ of the relevant system state.
The system represented by the pair of equations modeling dynamics and measurement $(f,g)$ is observable when data $\yvt{t}$ determine the system state $\xvt{t}.$
Identifying a set of biomarkers to render a system observable is equivalent to selecting measurements with $g$ that maximize our ability to determine a biological state $\xvt{t}$ throughout time.

Biological systems are highly complex, with individual cells containing millions of proteins—a scale that surpasses the typical focus of systems and control theory applications, such as jet engines or communication networks \cite{del2015biomolecular}.
While observability has been extensively studied in mathematical biology \cite{anguelova2004nonlinear, anguelova2007observability, villaverde2019observability, villaverde2019full}, its application to biomarker detection represents a new frontier that must account for the noisy, sparse, and high dimensional data in biology.
Metabolic and gene regulatory networks have been analyzed using structural observability, an approach that prioritizes scalability ($n$) over precision ($\theta_f$ and $\theta_g$) due to limited consideration of parameters learned from biological data \cite{lin1974structural, liu2013observability, liu2011controllability}.
Using time series transcriptomics data, Hasnain \textit{et al.} employed data-driven modeling and observability optimization to learn dynamics ($f$) and design biomakers ($g$) for pesticide detection \cite{hasnain2023learning}.
Still, many practicalities of biological systems and data remain unaddressed by observability theory.

\begin{figure*}[t]
    \centering
    \includegraphics[width=\linewidth]{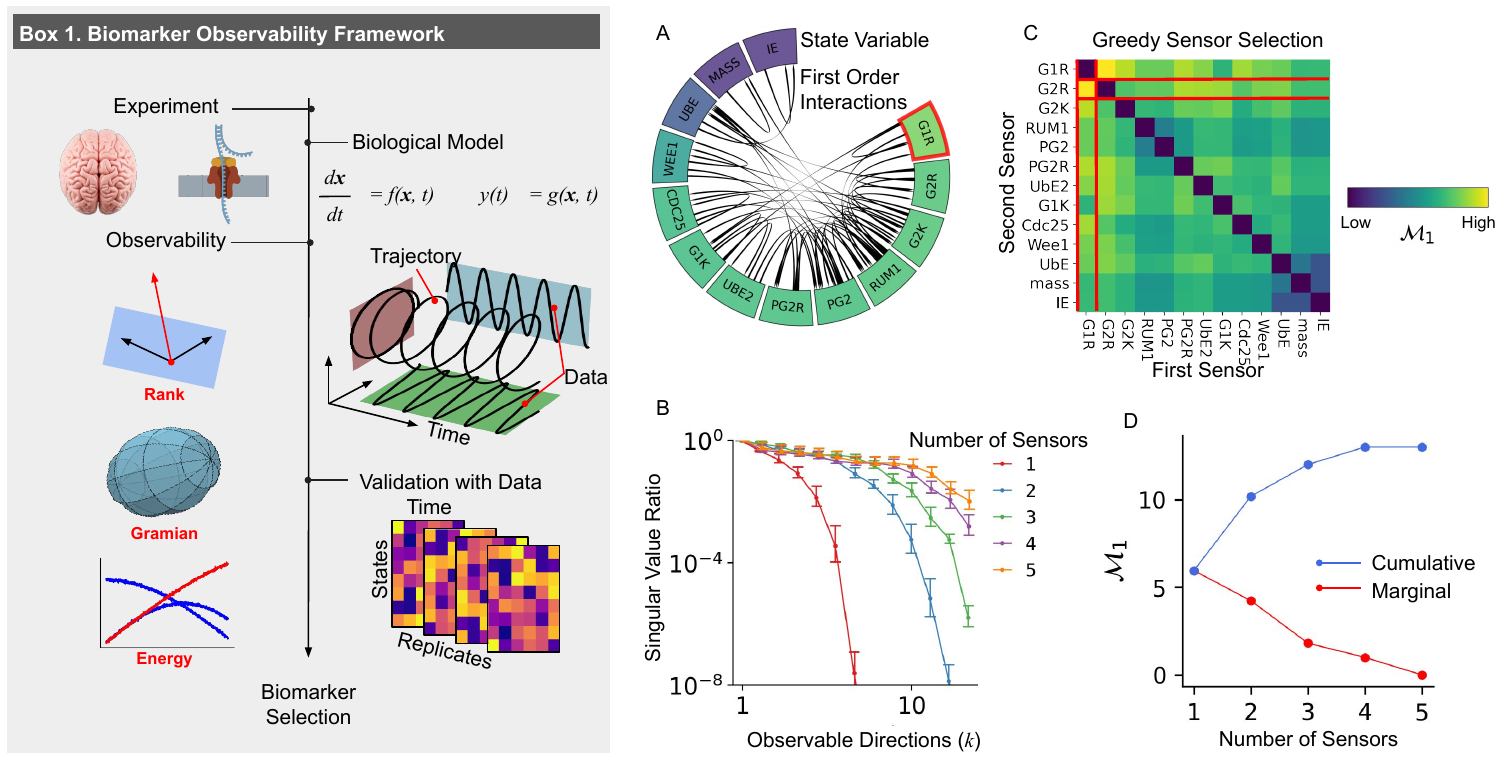}
    \vspace{-10mm}
    \caption{\small
    \textbf{Framework for applying observability to biological data.}
    \textbf{(Box 1)}
    Models of biological systems are constructed from experimental data, and sensor selection determines which low-dimensional representations of system trajectories will capture the most informative aspects of the system.
    \textbf{(A)} The thirteen state variables along with their first order interactions of the Tyson and Novak model are shown. Each state is colored according to their individual contribution to observability measured by $\Ms_1$ in the synthetic data. $G1R,$ which has the highest average contribution to observability, is boxed in red since it is selected as the first sensor by the greedy sensor selection algorithm.
    \textbf{(B)} The ratio of singular values $\sigma_k/\sigma_1$ of $\Os(\xv)$ measures $\Ms_1$ and increases with the number of sensors. This shows that when few sensors are used the smaller singular values are insignificant and the observability matrix $\Os(\xv)$ is approximately low rank.
    \textbf{(C)} The effective rank of $\Os(\xv)$ is shown when pairs of state variables are included in the sensor set. After \textit{G1R} is selected by the first iteration of the greedy algorithm, \textit{G2R} is the next best choice to maximize $\Ms_1.$
    \textbf{(D)} The observability $\Ms_1$ is shown over multiple iterations of the greedy algorithm. At each iteration, the observability increases and the contribution of the next sensor diminishes.
    }
    \label{fig1}
\end{figure*}

Biomarker discovery has previously relied on two advantages of biological systems: (1) a wealth of domain knowledge that exists independently of mathematical models ($f$ and $g$), and (2) an increasing array of high-dimensional, multimodal data and experimental techniques.
To bridge the gap between observability theory and biomarker discovery, we integrate observability into the biomarker detection problem as follows (\cref{fig1}):

\color{black}
\begin{enumerate}
    \item \textbf{Data Driven Biological Models:} 
    We apply techniques from Dynamic Mode Decomposition (DMD) and Data-Guided Control (DGC) to construct time-dependent models of gene expression.
    \item \textbf{Observability Analysis:}
    We present several measures of observability ($\Ms$, \cref{tab: solutions}) and provide optimization strategies. Dynamic Sensor Selection (DSS) methods are developed to reallocate sensors and optimize observability throughout time.
    \item \textbf{Biological Validation:}
    Observability-guided biomarkers are validated against established biological knowledge. Additionally, we incorporate chromosome conformation and other biological data as constraints to refine the observability analysis, ensuring alignment with biological priors.  
\end{enumerate}
Our work provides insights into the relationship between observability and the monitoring of biological systems, alongside the development of methods for DSS.
We introduce approaches for DSS and propose strategies to integrate gene expression (RNA-seq) and chromosome structure (Hi-C) data within the observability framework.
We demonstrate the utility and versatility of our framework across multiple datasets from genomics and beyond.

\section{Results}
%
The rank of the observability matrix $\Os(\xv)$ is a traditional criterion used to assess observability.
A system is observable when  $rank(\Os(\xv))=n$.
For the pair $(f,g)$ the observability matrix is:
\begin{equation}\label{eq: nom}
    \quad\Os(\xv)=\begin{bmatrix}
        \Ls_fg(\xv)\\
        \Ls^2_fg(\xv)\\
        \Ls^3_fg(\xv)\\
        \vdots
    \end{bmatrix}.
\end{equation}
Here, $\Ls_fg$ denotes the Lie derivative \cite{bloch2015nonholonomic}.
When $(f,g)$ are nonlinear, the matrix $\Os(\xv)$ depends on a particular state vector $\xv$, meaning observability is a local property that determines whether the system is locally observable at $\xv$ \cite{hermankrener}.
When the system is linear time-invariant (LTI), i.e. $f(\xv)=\Av\xv$ and $g(\xv)=\Cv\xv$ where $\Av$ and $\Cv$ are matrices, the observability matrix is $\Os=\begin{bmatrix}
    \Cv^\top&(\Cv\Av)^\top&(\Cv\Av^2)^\top&\dots
\end{bmatrix}^\top,$ and the rank criterion is the famous Kalman condition, which establishes a global observability property for all $\xv$ \cite{kalman1963mathematical}.

The application of rank-based criteria is impractical for high-dimensional systems with imperfect models of dynamics, as is often the case with biological systems.
Take, for example, the DNA replication model in fission yeast proposed by Novak and Tyson (\cref{fig1}A)) \cite{novak1997modeling}.
Their model comprises a differential equation with twelve state variables representing gene expression and one representing cell mass, forming a system that becomes observable if any gene is monitored (\textit{supplementary information} {\cref{eq: novak}}) \cite{liu2013observability}.
Yet, in practice, due to poor conditioning the observability matrix $\Os(\xv)$ is approximately low rank.
To demonstrate this, we used synthetic data to construct $\Os(\xv)$ for one thousand randomly selected state vectors $\xv$, testing each configuration where a single variable serves as the sensor.
Although $\Os(\xv)$ is full rank when using symbolic calculations, the singular values $\sigma_1,\dots,\sigma_{13}$ of $\Os(\xv)$ reveal that these matrices are only effectively low rank across all thousand simulated data points (\textit{supplementary information} \cref{fig: cc effective ranks}).
The simulated data shows that poor conditioning of the observability matrix $\Os(\xv)$ — characterized by the ratio $\sigma_k/\sigma_1$ being extremely small for $k=2,\dots,13$ — gives the appearance that $\Os(\xv) \neq n$, as if the system is locally unobservable at all sampled state vectors when only one sensor is utilize (\cref{fig1}B).
To address the practical concerns of rank based observability tests, a range of graded observability measures have been developed.

\paragraph{Observability Metrics.}
Beyond the Kalman condition, several metrics $\Ms$ to quantify observability have been proposed \cite{pasqualetti2014controllability}.
The Kalman condition, for instance, can be relaxed to measure the rank of the observability matrix:
\begin{align*}
    \Ms_1&=rank(\Os(\xv)) &\text{(directions).}
\end{align*}
When the system is not fully observable, i.e. $\Ms_1<n,$ the number of directions or principal components of the system that can be observed is given by $\Ms_1$ \cite{moore1981principal}.

The observability Gramian $\Gv_o$ is an alternative matrix to quantify observability.
For LTI systems, the observability Gramian $\Gv_o$ is defined as the solution to the Lyaponov equations:
\begin{equation*}
\begin{split}
    \Av^\top\Gv_o+\Gv_o\Av&=-\Cv^\top\Cv\quad\text{(continuous time),}\\
    \Av^\top\Gv_o\Av-\Gv_o&=-\Cv^\top\Cv\quad\text{(discrete time).}
\end{split}
\end{equation*}
This matrix facilitates the calculation of two additional observability metrics:
\begin{align*}
    \Ms_2 & = \xvt{0}^\top\Gv_o\xvt{0}&\text{(energy) }\label{eq: measures of observability}\\
    \Ms_3 & = trace(\Gv_o)&\text{(visibility)}.
\end{align*}
The energy metric $\Ms_2$ reflects the amplitude of the measured data $\yvt{t}$, and the visibility metric $\Ms_3$ is like an average measure of observability for each direction in the state space \cite{cortesi2014submodularity}.

The ability to compute the observability Gramian $\Gv_o$ and the observability matrix $\Os$ using various algorithms makes $\Ms_1$, $\Ms_2$, and $\Ms_3$ robust metrics well-suited for DSS.
Additional observability metrics include: (1) structural observability, which is favored for its scalability \cite{lin1974structural}, (2) algebraic observability, which is applicable to nonlinear systems \cite{sedoglavic2001probabilistic, liu2013observability}, and (3) additional Gramian based metrics \cite{pasqualetti2014controllability, cortesi2014submodularity} (\cref{tab: solutions}).

\begin{table*}[hb]
    \centering
    \begin{tabular}{clc|ccc|c}
        \toprule
        &Measure&&LTI&LTV&Nonlinear&DSS\\
        \hline
        $\Ms_1$&$rank(\Os)$&$\R$&\checkmark&\checkmark&\checkmark&\checkmark\\
        $\Ms_2$&Energy&$\R$&\checkmark&\checkmark&\checkmark&\checkmark\\
        $\Ms_3$&$trace(\Gv_o)$&$\R$&\checkmark&\checkmark&\checkmark&\checkmark\\

        $\Ms_4$&Algebraic \cite{diop1991nonlinear, sedoglavic2001probabilistic}&$0/1$&\checkmark&\xmark&\checkmark&\xmark\\
        $\Ms_5$&Structural \cite{lin1974structural}&$0/1$&\checkmark&\checkmark&\xmark&\xmark\\
        \bottomrule
    \end{tabular}
    \caption{\small \textbf{Observability Measures.} A comparison of five observability criteria, highlighting their condition (graded or binary), applicable dynamics, and compatibility with dynamic sensor selection. Algebraic observability is not suitable for time varying systems because the corresponding differential algebraic conditions require the system and its parameters remain constant \cite{sedoglavic2001probabilistic}. Structural observability, which stems from structural controllability, represents the first order interactions (i.e. the Jacobian of $f(\cdot)$) as a network. While the observability of this network indicates a linearization of the corresponding nonlinear system is observable, structural observability of the network does not guarantee the observability of the nonlinear systems \cite{diop1991nonlinear}. DSS is designed for nonlinear biological systems whose dynamics and sensors can be reallocated throughout time, which excludes the further consideration of $\Ms_4$ and $\Ms_5.$
    }
    \label{tab: solutions}
\end{table*}

\paragraph{Sensor Selection Problem.}
In order to maximize observability, the biomarker or sensor selection problem is formulated as:
\begin{equation}\label{eq: observability maximization}
    \max_{\text{sensors}} \quad \Ms 
    \quad \text{subject to experimental constraints.}
\end{equation}
Common constraints include a budget or inability to measure certain variables.
Since each variable may or may not be measured, there are $2^n$ candidate solutions to the sensor selection problem making the optimization challenging for high dimensional biological systems where $n$ is large.

To optimize observability of the Novak and Tyson model, we applied a greedy sensor selection algorithm (\textit{supplementary information} \cref{alg:greedy}).
The greedy approach selects sensors iteratively, at each step selecting the candidate sensor that maximize $\Ms_1$, averaged across all instances in the synthetic dataset.
\textit{G1R} was chosen first because it provided the highest rank of $\Os(\xv)$ (\cref{fig1}A).
With \textit{G1R} selected, the algorithm then evaluated the observability $\Ms_1$ when monitoring \textit{G1R} and each remaining state variable, selecting \textit{G2R} as the next sensor (\cref{fig1}C).
This process was repeated until $\Ms_1$ is maximized at 13, when monitoring the five variables: \textit{G1R}, \textit{G2R}, \textit{PG2R}, \textit{UbE}, and \textit{mass}.

When there is no limit on how much data can be measured in a simulation or experiment, a greedy algorithm finds the optimal solution to the sensor selection problem of \cref{eq: observability maximization}.
As the greedy algorithm iteratively selects sensors, the system's observability improves with each step.
This is reflected in both the ratio between singular values $\sigma_k/\sigma_1$ (\cref{fig1}B) and the average effective rank of $\Ms_1$ (\cref{fig1}D).
However, the marginal utility of each additional sensor decreases as fewer unobserved directions remain, reflecting diminishing returns in observability provided by each additional sensor.
The poor conditioning of $\Os(\xv)$ and the diminishing return from from the use of additional sensors leads us to our first result: practical application requires consideration of data quality, local system states, and resource constraints to ensure the effective use of biomarkers that theoretically make a system observable.

\paragraph{Biomarker Observability Depends on Biological State.}
Biomarker selection has traditionally relied on domain expertise to identify markers associated with specific biological states \cite{ray2010statistical}.
For example, the PIP-FUCCI system, a live cell microscopy approach for determining cell cycle stages (G1, S, G2, M), was developed based on prior knowledge of cell cycle dynamics and leverages biomarkers that vary with cell state \cite{grant2018accurate, zielke2015fucci}.
Imaged expression data from three genes reveals cell cycle stages and transitions: \textit{CDT1} (G1), \textit{PCNA} (S), and \textit{GEM} (S/G2/M).
Because cells progress through the cycle at different rates — affected by cell type, experimental conditions, and cell-to-cell variability — the position within the cell cycle determines the relevance of the PIP-FUCCI biomarkers at different points in time \cite{darzynkiewicz1982cell, spiller2010measurement}.

\begin{figure}[H]
    \centering
    \includegraphics[width=\linewidth]{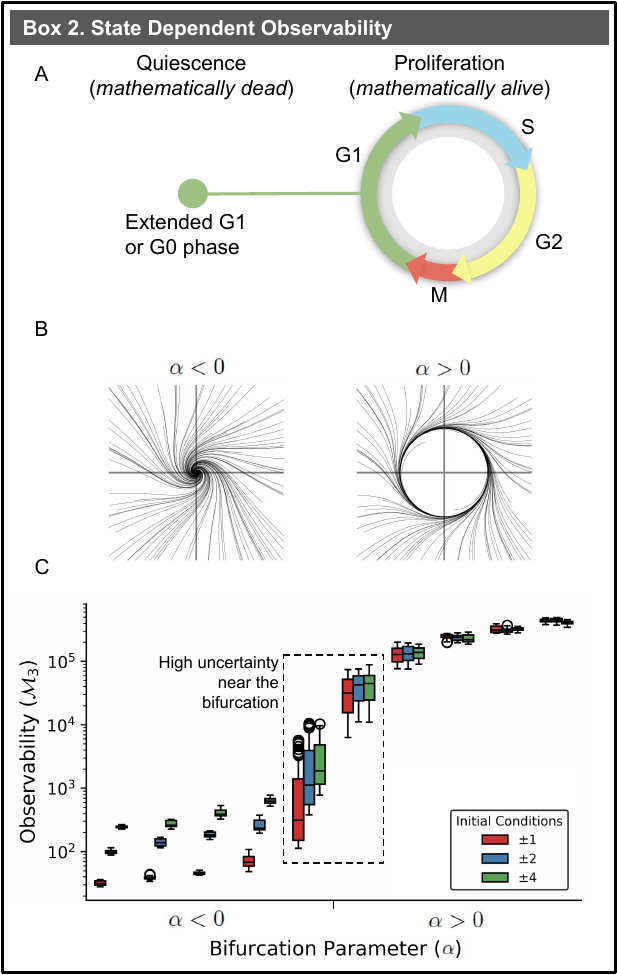}
    \vspace{-7mm}
    \caption{\small
    \textbf{State-Dependent Observability.}
    \textbf{(A)} A cell's progression through the cell cycle—whether transitioning through phases during proliferation or stalling in G1/G0 during quiescence—is mediated by CDK2 activity \cite{spencer2013proliferation}.
    \textbf{(B)} The Andronov-Hopf oscillator demonstrates either asymptotically stable or periodic limit cycle behavior, depending on the parameter \( \alpha \).
    \textbf{(C)} The transition from stable to periodic behavior in the Andronov-Hopf oscillator coincides with an increase in observability. Initial conditions used to construct the empirical observability Gramians were selected by sampling $x_1$ and $x_2$ from uniform distributions bounded by $\pm 1,\ \pm2,$ and $\pm 4.$
    }
    \label{fig:EMG}
\end{figure}

As cells progress through the cell cycle, they may stall in the G1 phase, often referred to as G0, entering a state known as quiescence (\cref{fig:EMG}A).
During quiescence, the cell ceases to divide, much like a system reaching a stable equilibrium.
Quiescent cancer cell are linked to high cancer recurrence, as these cells can re-enter the cell cycle, and the reduced cell cycle activity diminishes the effectiveness of chemo- and immunotherapies \cite{lindell2023quiescent, li2011stem, chen2016cancer, lee2020regulatory}.
The proliferation-quiesence bifurcation is described mathematical as a transition between stability and periodicity in dynamical systems and coincides with a shift in observability \cite{riba2022cell}.
Smale's two-cell system illustrates this concept, exhibiting stable solutions that correspond to quiescence and periodic solutions that resemble progression through the cell cycle \cite{smale1976mathematical}.
As a special case of Turing's equations of morphogenesis, these dynamics have been characterized as “\textit{mathematically dead}” when stable and “\textit{mathematically alive}” when oscillatory \cite{turing1952chemical, chua2005local}.

Systems that are \textit{mathematically alive} exhibit heightened observability.
The Hopf bifurcation, the archetypal example of the transition between stable and periodic dynamics \cite{marsden2012hopf}, is observed in the Andronov-Hopf oscillator:
\begin{equation}\label{eq: Andronov-hopf}
    \begin{split}
            \frac{d x_1}{dt} &= \alpha x_1 - x_2 - x_1 \left(x_1^2 + x_2^2\right) \\
            \frac{dx_2}{dt} &= x_1 + \alpha x_2 - x_2 \left(x_1^2 + x_2^2\right).\\
    \end{split}
\end{equation}
As the parameter $\alpha$ transitions from negative to positive values (\cref{fig:EMG}B), the system shifts from being \textit{mathematically dead} to \textit{alive!}
To assess how observability transitions between the \textit{mathematically dead} and \textit{alive} states, we select $x_1$ as a sensor and measure the output with $y=x_1.$
With the dynamics of the Andronov-Hopf oscillator and fixed sensor, empirical observability Gramians were constructed from simulated data with both \textit{dead} and \textit{alive} choices of $\alpha$ \cite{himpe2018emgr}.
Empirical observability Gramians $\Gv_o$ are constructed from simulations and perturbations of the system \cref{eq: Andronov-hopf}, as opposed to solving the Lyapunov equations or other approximation techniques (\textit{supplementary information} \S\ref{SI: empirical gramian}).
From the observability Gramian for each simulation, $\Ms_3$ measured the observability of the system (\cref{fig:EMG}C).
The degree of observability of this system is primarily controlled by the bifurcation parameter $\alpha$ and modulated by the choice of initial condition for the simulation. 
Variance in
\noindent
observability diminishes for $\alpha>0$ because simulated trajectories are driven toward similar limit cycles or cyclic patterns.
In contrast, when the system is stable and $\alpha < 0$, observability varies significantly between simulations with different initial conditions.
This occurs because the starting point becomes the key differentiator for how quickly trajectories converge to the stable point.
In other words, \textit{mathematically alive} systems are more observable because the periodic behavior causes the system state cover a wider range of the state space, which increases the measurable information in the sensor data.

The variation in observability governed by $\alpha$ serves as a mathematical analog to insights from PIP-FUCCI biomarkers: the observability provided by sensors is not constant.
Although the gene \textit{CDT1} is a biomarker for the G1 phase, its contribution to observability diminishes during other cell cycle phases, necessitating the use of \textit{PCNA} and \textit{GEM} to delineate each phase.
Similarly, as $\alpha$ increases, the contribution of the sensor $x_1$ to system observability increases by a factor of $10^3$.
The highest variability in observability occurs during the transition between periodic and stable behaviors of the system, where $\alpha$ is near 0, at the bifurcation occurs.
Thus, the utility of measuring $y=x_1$ is least certain at the point where the system is nearest to transitioning between stable and periodic behaviors.
This observation leads us to our second main result: the dependence of biomarkers on biological state requires time-dependent biomarker selection.

\paragraph{Dynamic Sensor Selection (DSS).}
%
To address the need to identify and allocate biomarkers over shifts in the underlying dynamics, we developed DSS to identify time varying biomarkers based upon models of dynamics learned from time series data genomics.
Due to the high dimensionality and relatively few time points found in genomics data, this approach is tailored to discrete time LTI and linear time-varying (LTV) models of dynamics that can be learned from time series transcriptomics data (\textit{supplementary information} \S\ref{SI: learning dynamics}).
After learning a model of dynamics $\xvt{t+1}=\Avt{t}\xvt{t},$ DSS adapts the sensor selection problem (\cref{eq: observability maximization}) to select sensors for each point in time:
\begin{equation}\label{eq: DSS observability maximization}
    \max_{\text{sensors}(t)} \quad \Ms 
    \quad \text{subject to constraints at time $t$.}
\end{equation}
The selected sensors are placed in a measurement matrix $\Cvt{t}$ that varies with time.
While time-dependent systems have been studied in the context of controllability and robustness \cite{silverman1967controllability, sastry1982robustness}, sensor selection for time-dependent systems has only received limited attention in the literature.

To optimize \cref{eq: DSS observability maximization} for $\Ms_1$, $\Ms_2$, and $\Ms_3$, the time-dependent observability Gramian is required.
The observability Gramian of a LTV system from time $t_0$ to time $t$ is:
\begin{equation}\label{eq: observability gramian}
    \Gv_o(t_0, t) = \sum_{k=t_0}^t \Phi(t_0,k)^\top \Cvt{k}^\top\Cvt{k}\Phi(t_0, k),
\end{equation}
where $\Phi(t_0, t)$ is the time-dependent state transition matrix from $t_0$ to $t$, given by:
\begin{equation*}
    \Phi(t_0, t) = \Avt{t} \cdots \Avt{t_0+1} \Avt{t_0}.
\end{equation*}

Maximization of $\Ms_2$ is achieved by solving the eigenvalue problem:
\begin{equation}\label{eq: eig problem C(t)}
    \Gv(t,t_0)\Cvt{t}^\top=\Cvt{t}^\top\Dv,
\end{equation}
where $\Gv(t, t_0)$ is a Gram matrix learned from data as 
\begin{equation*}
    \Gv(t, t_0)= \Phi(t,t_0)\xvt{0}\xvt{0}^\top\Phi(t,t_0)^\top.
\end{equation*}
The columns of $\Cvt{t}$ in \cref{eq: eig problem C(t)} correspond to optimal sensor placement at time $t$ and the contribution to observability is weighted by the eigenvalues found in $\Dv$.

Measure $\Ms_3$ can be maximized with a linear program.
The LTV observability Gramian in \cref{eq: observability gramian} can be expressed equivalently as:
\begin{equation}
    \Gv_o(t_0,t)=\sum_{i=t_0}^t\sum_{j=1}^n\delta_{ij}\Wv_{ij},
\end{equation}
where $\Wv_{ij}=\Phi(t_0,i)^\top (\cv_{ij})^\top(\cv_{ij})\Phi(t_0, i)$, $\delta_{ij}$ indicates 1 if variable $j$ is measured at time $i$ and 0 otherwise, and $\cv_{ij}$ is a row vector with the $j$th entry as 1 and 0 otherwise.
Because the matrix trace is linear, when $\delta_{ij}$ is relaxed to a continuous value, i.e. $0\leq \delta_{ij}\leq 1,$ a linear program can solve the optimization:
\begin{equation}\label{eq: linear program}
    \max_{\delta}\Ms_3\quad\text{ where } 0\leq\delta_{ij}\leq 1.
\end{equation}
Constraints can be incorporated into this optimization, such as restricting the number of sensors $p$ at time $t$ with the constraint: $\sum_{j=1}^{n}\delta_{ij}\leq p(t).$
See \textit{supplementary information} \S\ref{SI: DSS optimizations} for precise details on these optimizations.
In the following sections, we demonstrate a range of applications for the DSS methodology in biological systems.

\paragraph{Estimating Unmeasured Genes.}
%
To evaluate DSS, we identified genes critical to observing the dynamics of \textit{Pseudomonas fluorescens} SBW25, a bacterium used for insect control, from data collected by Hasnain \textit{et al.} \cite{hasnain2023learning}.
The dataset comprises time series spanning nine time points and 600 genes, representing a high-dimensional system that presents challenges for observation.
Gene regulation in the bacteria or other cells is represented by \cref{eq: dynamics}, where $\xvt{t}_i$ denotes the expression of the $i$th gene, and $\Avt{t}$ is the gene regulatory network at time $t$.
The matrix element $\Avt{t}_{ij}$ specifies the influence of gene $j$ on gene $i$ at time point $t$.
We selected two sets of sensors that (1) optimize $\Ms_2$ over time with DSS and (2) optimize $\Ms_2$ with fixed sensors throughout the experiment.

\begin{figure*}[t]
    \includegraphics[width=\linewidth]{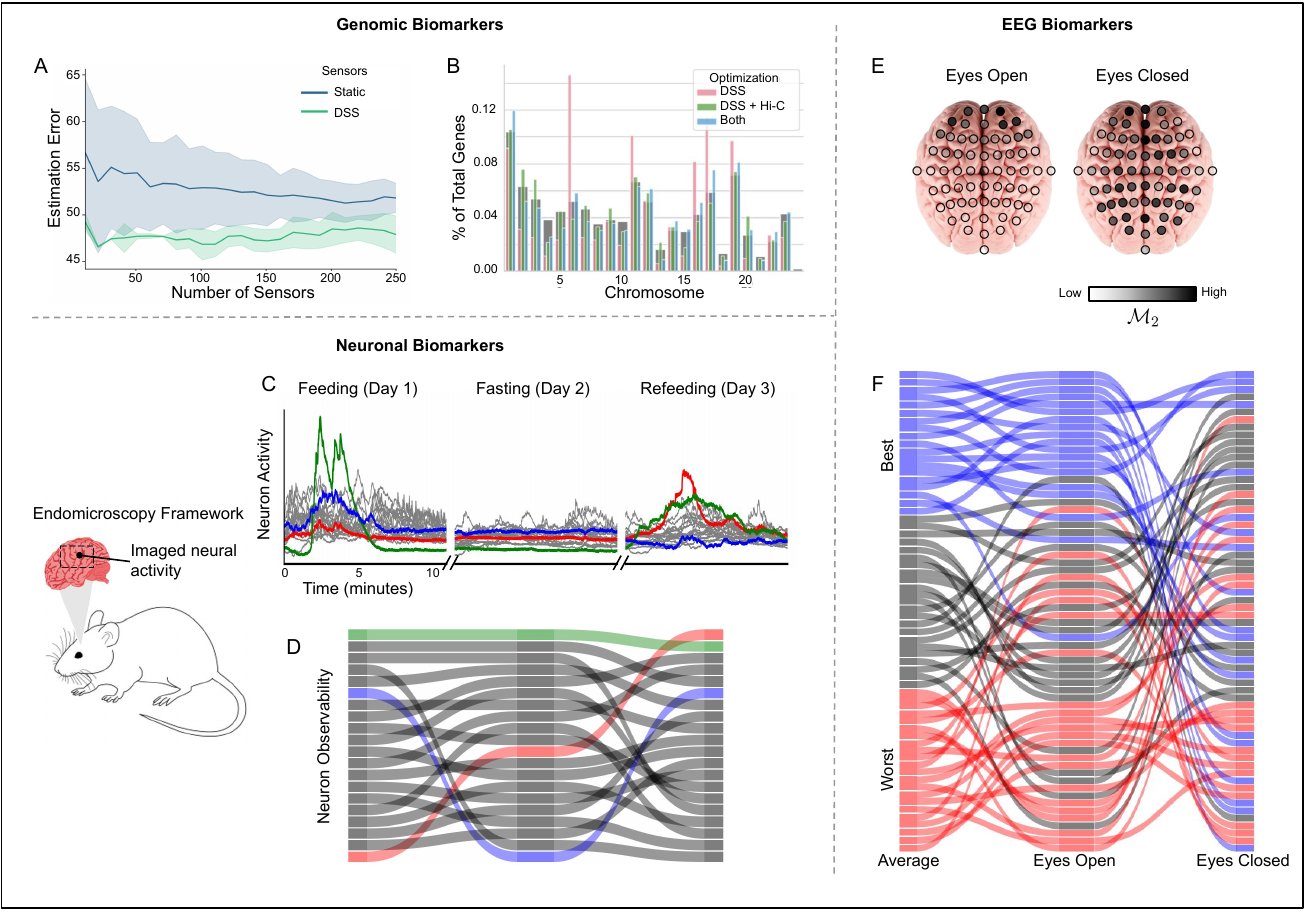}
    \vspace{-5mm}
    \caption{\small
    \textbf{Biomarker Selection from Time Series Data.}
    \textbf{(A)} DSS improves the estimation error of individual genes from biomarker data relative to the use of biomarkers that are fixed throughout time.
    \textbf{(B)} Constraining the sensor selection problem with Hi-C positions highly observable biomarker genes on chromosomes to more closely reflect the spatial distribution of genes within the nucleus, as indicated by the gray background. The positions of the top 10\% of biomarkers selected with unconstrained DSS, DSS constrained by Hi-C data, and biomarkers common to both methods are shown in pink, green, and blue, respectively.
    \textbf{(C)} The time series neuron activity was collected for 10 minute segments on three consecutive days. The recorded activity extracted from twenty neurons is shown, with the activity of three neurons highlighted in red, green, and blue.
    \textbf{(D)} Throughout the three day period, the observability contributed by each neuron varies greatly. The neuron indicated in red, which initially is the worst sensors, becomes the most observable as its overall activity becomes the largest in day 3.
    \textbf{(E)} The spatial position of 64 EEG leads colored by their contribution to observability.
    \textbf{(F)} The signals from each of the 64 EEG leads are ranked based on their observability, with the average rank representing each sensor's mean ranking across all six tasks.
    }
    \label{fig:brain}
\end{figure*}

We found that DSS improves state estimation relative to using fixed sensors.
To test estimation capabilities, after selecting biomarkers, the data is divided into observable biomarker and unobservable non-biomarker datasets.
Estimating the non-biomarker gene expression can then be formulated as solving the following least squares problem:
\begin{equation}\label{eq: estimation} 
    \min_{\hat{\xv}}\|\Yv-\Os\hat{\xv}\|,
\end{equation}
where $\hat{\xv}$ is an efficient estimator of $\xvt{0}$ \cite{sorenson1970least}.
The biomarker data $\yvt{t}$ is assembled in a matrix $\Yv=\begin{bmatrix}
    \yvt{0}^\top&\yvt{1}^\top&\dots
\end{bmatrix}^\top.$
The estimation of unmeasured genes in \cref{eq: estimation} has the solution $\hat{\xv}=\Os^\dagger\Yv,$ where $\dagger$ denotes the pseudoinverse \cite{barata2012moore}.
To measure the quality of the estimation of $\hat{\xvt{0}},$ it is compared to the true data of $\xvt{0}.$
This error can be measured using several metrics, e.g. $\|\hat{\xv}-\xvt{0}\|$, but the component-wise error is most relevant to assessing the error of estimating individual genes.
DSS consistently improved the median estimation error for each of the 600 genes, regardless of the number of sensors used (\cref{fig:brain}A).
The median estimation error measures the ability to estimate the expression values of individual genes.
In spite of the biomarkers and system not satisfying the Kalman rank condition or the Popov-Belevitch-Hautus test, the incorporation of time-varying dynamics and sensors enables unmeasured genes to be estimated with an error within 50\%, a level of accuracy that is practically useful for many real-world applications.

\paragraph{Functional Observers for Cellular Reprogramming.}
While DSS enhances state estimation of unmeasured missing gene expression values, many biomedical applications rely on biomarkers to indicate phenotypes or cell types.
Early detection of cellular reprogramming, a process that transforms cell type and induces a shift in the dynamics of the cell's transcriptional program, is an important and unresolved challenge in biomanufacturing \cite{yamanaka2009elite}.
This task falls under the framework of functional observability, where the goal is to select biomarkers or sensors that enable the estimation of specific modes of the unmeasured states — such as phenotype — without reconstruction of all unmeasured state variables or genes \cite{trinh2011functional}.

A system is functionally observable for the modes defined by the rows of the matrix $\Fv$ if
\begin{equation}
    {rank}\begin{bmatrix}
        \Os\\
        \Fv
    \end{bmatrix}={rank} (\Os).
\end{equation}
Here, $\Fv$ is a matrix where each row represents a functionally observable mode or direction of the system.
When $\Fv$ is sparse, with only one nonzero entry per row, individual states can be estimated — a property known as targeted observability \cite{montanari2022functional, zhang2024functional}.
In contrast, when the rows of $\Fv$ are dense, each row can represent a cell type, and each element within a row can correspond to the expression of a particular gene for that cell type, enabling the system to be functionally observable for a biological state.

A system is always functionally observable for the principal components (or right singular vectors) of $\Os$, which are utilized in the estimation framework of \cref{eq: estimation} \cite{moore1981principal}.
By applying the Singular Value Decomposition (SVD) to the observability matrix $\Os=\Uv\Sigma\Vv^\top,$ the pseudoinverse $\Os^\dagger$ is expressed as $\Os^\dagger=\Vv\Sigma\inv\Uv^\top.$
Consequently, the state estimation of $\hat{\xv}$ is computed as:
\begin{equation}
\begin{split}
    \hat{\xv}&=\Os^\dagger\Yv=\Vv\Sigma^{-1}\Uv^\top\Yv.
\end{split}
\end{equation}
This equation demonstrates that the state estimate $\hat{\xv}$ is a linear combination of the right singular vectors ($\Vv$) weighted by the contributions of the nonzero singular values ($\Sigma^{-1}$) and the data ($\Uv^\top\Yv$).
Since the rows of $\Vv^\top$ associated with nonzero singular values are in the row space of $\Os,$ the system is functionally observable for the modes described by $\Fv=\Vv^\top.$

With time series gene expression data from a recreation of Weintraub's seminal 1989 reprogramming experiment, we built a functional observer for the reprogramming of fibroblasts to skeletal muscle \cite{weintraub1989activation}.
To monitor the progression of myogenic reprogramming initiated by the introduction of the transcription factor \textit{MYOD}, bulk RNA-seq data was collected at 8-hour intervals \cite{liu2018genome}.
Cellular reprogramming remains characterized by partial reprogramming and low efficiency, which result in weak and noisy signals.
To address this, we developed two LTV models of gene expression dynamics to amplify the reprogramming signal and facilitate observer construction.
The first model (Model 1) encompassed all 19,235 genes measured during the experiment, while the second model (Model 2) focused on 406 genes involved in cell cycle regulation and myogenic lineages, aiming to enhance the weak reprogramming signal.
To identify biomarkers, we optimized $\Ms_2$ on Model 2 to identify which reprogramming genes most strongly contribute to system observability.
The top ranked reprogramming biomarkers identified by Model 2 were used to monitor Model 1.

To identify the functionally observable modes of Model 1 with the selected biomarkers, we constructed the observability matrix $\Os$ of Model 1, performed the SVD $\Os=\Uv\Sigma\Vv^\top,$ and considered the rows of $\Vv^\top$ associated with nonzero singular values.
Each row of $\Vv^\top$ is a functionally observable mode, and the entries of a row in $\Vv^\top$ indicate the expression of different genes.
We performed cell type and functional enrichment to determine the biological processes that could be observed associated with each mode.
This revealed cell types such as fibroblasts, myofibroblasts, and myoblasts, all known to be involved in myogenic reprogramming (\textit{supplemental information} \cref{fig: cell type enrichment}).
Similarly, there is a strong preference for myogenic and cell cycle genes, both involved in reprogramming, to be heavily weighted in the functionally observable modes.
Further enrichment analysis highlights cellular activities like the defense response to viruses (\textproc{GO:0051607}), aligning with the expected response due to Lentiviral reprogramming, and regulation of the cell cycle (\textproc{GO:0051726}) and smooth muscle cell proliferation (\textproc{GO:0048661}), consistent with the cell division and differentiation that occurred in these data (\textit{supplemental information} \cref{fig: pathway enrichment}).
The enrichment of functionally observable modes corresponding to biological states and processes consistent with the experiment suggests that, despite the low reprogramming signal and substantial noise, biomarkers identified using DSS are well-suited for monitoring cellular reprogramming.

\paragraph{Chromatin Informed Biomarkers.}
Integrating biological insights or domain knowledge that is not captured in the system model $f(\cdot)$ or state space $\xv$ into the observability-guided biomarker selection framework helps align the sensor selection problem with practical biological considerations.
For instance, monitoring multiple biomarkers within a transcription factory -- a group of genes that is colocalized in the nucleus where the genes are often coregulated -- may provide redundant information \cite{rieder2012transcription, chen2015functional}.
Modifying the sensor selection problem to limit the number of biomarkers per transcription factory or satisfy other requirements is achieved by modifying the constraints of \cref{eq: DSS observability maximization}.

To select biomarkers in the context of transcription factors, we use Hi-C data, which provides information on genome structure, to constrain \cref{eq: observability maximization}.
With Hi-C data from the study by \cite{liu2018genome}, we generated gene-by-gene Hi-C matrices representing observed contact frequencies between genes (\textit{supplementary information} \cref{fig: geneXgeneHiC}).
From the gene-centric Hi-C matrix, we performed hierarchical clustering and optimized the silhouette score to construct gene clusters $c_1, \dots, c_l$ where each cluster represents a group of genes that are likely proximal to one another (\textit{supplementary information} \cref{fig: hicGeneXgene 2}).
Then, we constrained \cref{eq: observability maximization} to prevent the simultaneous selection of multiple genes found within the same cluster:
\begin{equation*}
\setlength{\jot}{-10pt} 
\begin{split}
    \max_{\text{sensors}({t})} \Ms \quad&\text{such that at most one gene is measured}\\
    &\text{per cluster at each time $t$.}
\end{split}
\end{equation*}
To maximize $\Ms_2$, we applied a greedy heuristic by first solving the unconstrained maximization, then selecting the top ranked sensors that meet the constraints.
For $\Ms_3$, these constraints from Hi-C data can directly be incorporated into the linear program used to maximize \cref{eq: linear program}.

Although this constrained optimization yields a lower utility of the objective function $\Ms$, the selected biomarkers have two practical advantages.
First, they are distributed across the genome in a pattern that mirrors natural gene placement across chromosomes (\cref{fig:brain}B).
Second, their performance is comparable to that of biomarkers selected from the unconstrained dataset. 
Integrating chromatin-informed constraints into the biomarker selection process, ensures observability maximization in the context of and consistent with prior biological knowledge.

\paragraph{Beyond the Genome.}
To demonstrate the utility of observability guided biomarkers beyond genomics, we applied DSS to \textit{in vivo} single-cell endomicroscopic signals collected by \cite{sweeney2021network}.
We constructed a LTI model for each experimental phase — feeding, fasting, and refeeding — and identified the contribution to observability (\cref{fig:brain}C).
Our results revealed substantial shifts in monitoring utility across neurons.
Notably, the neuron with the lowest initial output energy ($\Ms_2$) exhibited the highest contribution to observability during refeeding (\cref{fig:brain}D).
A similar pattern is confirmed for $\Ms_3$.
The varying contributions to observability—where some neurons consistently act as good sensors, others are transiently effective only during feeding states, and some gain significant observability contributions after fasting—underscore the importance of dynamically selecting sensors.

At a larger scale, we applied DSS to assess the observability of electroencephalogram (EEG) signals \cite{schalk2004bci2000}.
Brain activity, sampled from 64 EEG leads at 160 Hz across a cohort of over 100 patients, provides high-quality time series data well suited for modeling and observability analysis \cite{pequito2015minimum, gupta2019learning}.
We evaluated the observability contribution of each of the 64 EEG leads as participants transitioned through various states, including open-eye and closed-eye conditions and four tasks involving hand and foot movements (\cref{fig:brain}E).
Notably, sensor performance varied significantly, particularly between the open-eye and closed-eye conditions (\cref{fig:brain}F, \textit{supplementary information} \cref{fig: full EEG data}).
These findings highlight the applicability of DSS and observability-guided biomarkers to a range of critical applications and data modalities.

\section{Discussion}
In this work we extended the tools of observability theory to identify biomarkers, accounting for the practicalities and constraints of experimental biological data.
Our key findings include (1) that the contribution to observability of biomarkers depends on the biological state.
This work establishes a connection between the state-dependent utility of biomarkers and the concepts of local and time-dependent observability.
Moreover, (2) we developed DSS as a mechanism to optimize observability as the dynamics of the underlying system change.
The development of DSS provides a computational approach for sensor selection in time-varying systems, a concept long recognized in observability theory but only recently made feasible for genomic biomarkers due to advances in measurement technologies.

Our application to real biomedical data demonstrates both the versatility of this approach and highlights practicalities of observability that are often overlooked in theoretical discussions and the absence of real data.
In particular, this work relaxes the need for mathematical modeling of biological systems by leveraging data-guided modeling, enabling the detection of biomarkers through observability analysis in any biomedical time series data. 
This also highlights the relevance of ensuring theoretical criteria can be validated numerically, within real or synthetic data.

This work also raises several research directions worthy of future pursuit.
First, while measured data can determine the unmeasured states of an observable system, there are many ways and algorithms to perform such an estimation.
Here, we have used the most basic approach, performing a least squares estimation; this leaves open several avenues for the design and construction of observers tailored to the unique noise, sparsity, and destructive nature of transcriptomics assays.
The development of observers is important for realizing the utility of observability-guided biomarkers with emerging sequencing technologies.
Second, the state space representation of a cell is crucial for determining observability criteria \cite{bunne2024build}.
In this work, we adopted the raw data as the state space, but this may not be the most optimal approach.
Future research could explore enhanced representations of a cell that integrate both genomic structure and function by incorporating additional data modalities.
Third, theoretical investigations into bifurcation observability, particularly in the context of Smale's two-cell system, are closely related to the bifurcation control problem and warrant further attention \cite{chen2000bifurcation}.
This areas, and others, must be further developed to leverage the rapid development of emerging experimental and computational biotechnologies.

\paragraph{Acknowledgments.} {We thank the members of the Rajapakse Lab for helpful and inspiring discussions. This work is supported by Air Force Office of Scientific Research (AFOSR) under award number FA9550-22-1-0215 (IR), FA9550-23-1-0400 (AB), NSF DMS-2103026 (AB) and NIGMS GM150581 (JP).}
\end{multicols}

\begin{multicols}{2}
\bibliography{refs2}
\end{multicols}

\newpage

\part*{Supplemental Information}
The supplemental information is divided into three sections: \S\ref{SI mm} contains description of the models, datasets, parameters, and methods used in the paper; \S\ref{SI: DSS optimizations} contains details regarding the DSS optimization methods, and \S\ref{SI: figures} contains supplemental figures referenced throughout the main text.

\section{Materials and Methods}\label{SI mm}

\subsection{DNA Replication Control}

\paragraph{Model Equations.}
\Cref{eq: novak} presents the model of DNA replication control originally introduced in \cite{novak1997modeling} and further analyzed in \cite{liu2013observability}.
We note that the dynamics considered in these two studies differ slightly, and we adopt the model and parameterization from \cite{liu2013observability} to ensure consistency with their observability analysis.
The variable $R$ in these equations denotes the $RUM1$ gene of the original model.
\begin{equation}\label{eq: novak}
\begin{split}
\frac{d Cdc25}{dt} &= - \frac{K_{CR} \cdot Cdc25}{K_{MCR} + Cdc25} + \frac{K_C \cdot Cdc25 \cdot (G2K + \beta \cdot PG2)}{K_{MC} + 1 - Cdc25} \\
\frac{d G1K}{dt} &= K_5 + (K_4 + K_{8R}) \cdot G1R - K_8 \cdot G1K \cdot R - G1K \cdot (V_{6P} \cdot (1 - UbE2) + V_6 \cdot UbE2) \\
\frac{d G1R}{dt} &= - K_4 \cdot G1R - K_{6P} \cdot G1R - K_{8R} \cdot G1R + K_8 \cdot G1K \cdot R \\
\frac{d G2K}{dt} &= K_1 + (K_4 + K_{7R}) \cdot G2R + (V_{25P} \cdot (1 - Cdc25) + V_{25} \cdot Cdc25) \cdot PG2 \\
&\quad - K_K \cdot G2K \cdot R - G2K \cdot (V_{2P} \cdot (1 - UbE) + V_2 \cdot UbE) - G2K \cdot (V_{WP} \cdot (1 - Wee1) + V_W \cdot Wee1) \\
\frac{d G2R}{dt} &= - K_4 \cdot G2R - K_{7R} \cdot G2R + K_7 \cdot G2K \cdot R - G2R \cdot (K_{2P} + V_{2P} \cdot (1 - UbE) + V_2 \cdot UbE) \\
\frac{d IE}{dt} &= - \frac{K_{IR} \cdot IE}{K_{MIR} + IE} + \frac{K_I \cdot IEC \cdot (G2K + \beta \cdot PG2)}{K_{MI} + IEC} \\
\frac{d \text{mass}}{dt} &= \mu \cdot \text{mass} \\
\frac{d PG2}{dt} &= - (V_{25P} \cdot (1 - Cdc25) + V_{25} \cdot Cdc25) \cdot PG2 + K_4 \cdot PG2R + K_{7R} \cdot PG2R \\
&\quad - K_7 \cdot PG2 \cdot R - PG2 \cdot (V_{2P} \cdot (1 - UbE) + V_2 \cdot UbE) + G2K \cdot (V_{WP} \cdot (1 - Wee1) + V_W \cdot Wee1) \\
\frac{d PG2R}{dt} &= - K_4 \cdot PG2R - K_{7R} \cdot PG2R + K_7 \cdot PG2 \cdot R - PG2R \cdot (K_{2P} + V_{2P} \cdot (1 - UbE) + V_2 \cdot UbE) \\
\frac{d R}{dt} &= K_3 + K_{6P} \cdot G1R + K_{8R} \cdot G1R + K_{7R} \cdot G2R + K_{7R} \cdot PG2R \\
&\quad - K_4 \cdot R - K_8 \cdot G1K \cdot R - K_7 \cdot G2K \cdot R - K_7 \cdot PG2 \cdot R \\
&\quad - \frac{K_P \cdot \text{mass} \cdot (CIG1 + \alpha \cdot G1K + G2K + \beta \cdot PG2 \cdot R)}{K_{MP} + R} + (G2R + PG2R) \cdot (K_{2P} + V_{2P} \cdot (1 - UbE) + V_2 \cdot UbE) \\
\frac{d UbE}{dt} &= - \frac{K_{UR} \cdot UbE}{K_{MUR} + UbE} + \frac{K_U \cdot IE \cdot (1 - UbE)}{K_{MU} + 1 - UbE} \\
\frac{d UbE2}{dt} &= - \frac{K_{UR2} \cdot UbE2}{K_{MUR2} + UbE2} + \frac{K_{U2} \cdot (G2K + \beta \cdot PG2) \cdot (1 - UbE2)}{K_{MU2} + 1 - UbE2} \\
\frac{d Wee1}{dt} &= - \frac{K_W \cdot (G2K + \beta \cdot PG2) + Wee1}{K_{WR} + Wee1} + \frac{K_{WR} \cdot (1 - Wee1)}{K_{MWR} + 1 - Wee1} \\
\end{split}
\end{equation}

\paragraph{Observability in the Synthetic Dataset.}
We used synthetic data to asses the observability of \cref{eq: novak} by validating the rank condition.
One thousand state vectors $\xv$ were randomly sampled, with each element chosen uniformly at random between 0 and 1.
For each state vector $\xv$ and for each possible sensor of the 13 state variables, the local, nonlinear observability matrix $\Os_i(\xv)$ was constructed, where $i$ denotes which of the state variables was used as the sensor.
To determine the rank of the matrix, the Singular Value Decomposition (SVD) was performed on each matrix $\Os_i(\xv)$ in order to obtain the singular values $\sigma_1\geq\dots\geq\sigma_{13}$.

\begin{figure}[h]
    \centering
    \includegraphics[width=\linewidth]{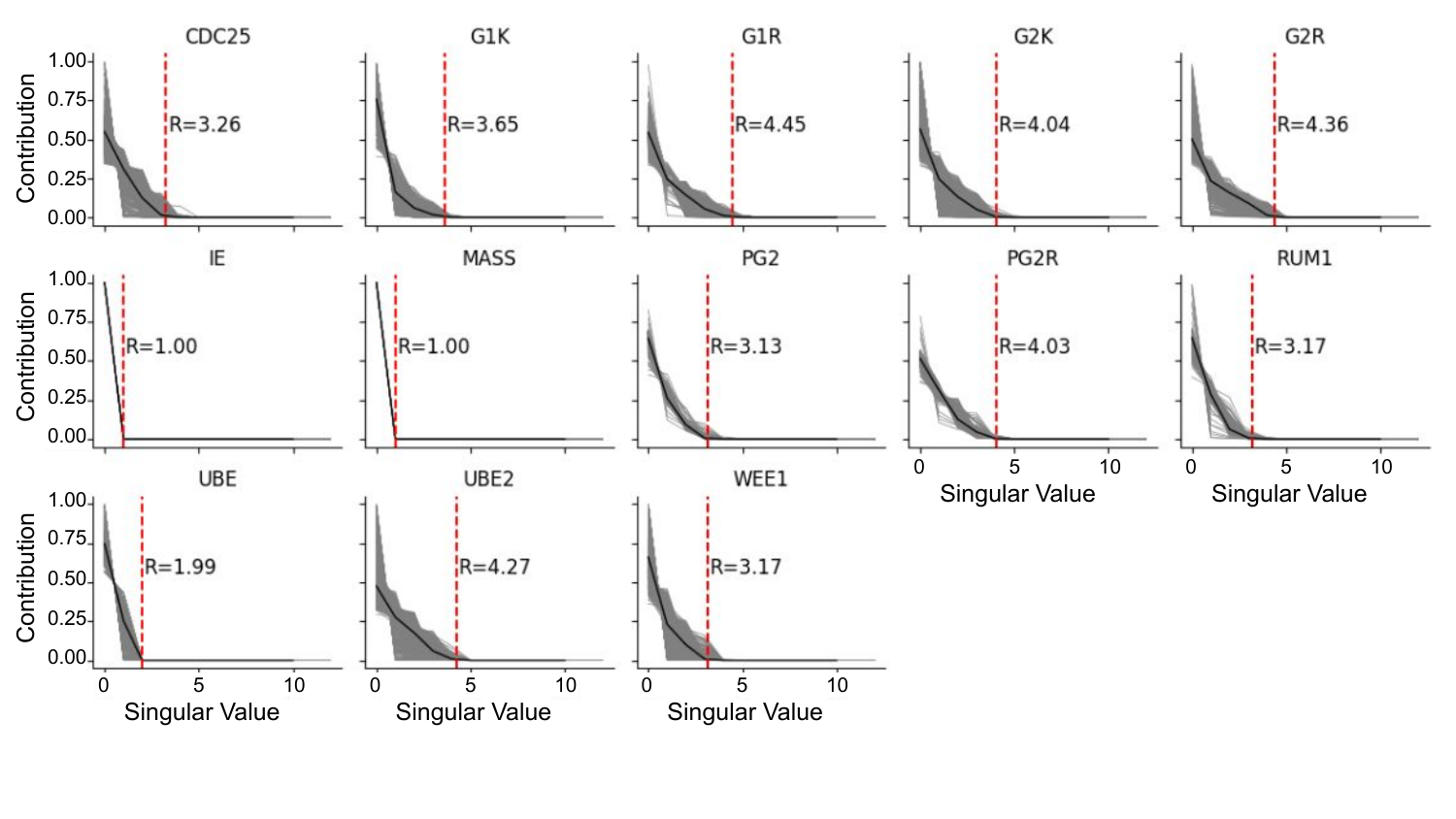}
    \vspace{-10mm}
    \caption{\textbf{Effective Rank of Observability Matrices.} The effective rank \textbf{R} of each sensor is shown using $\varepsilon=10^{-5}$. Although the system is observable according to Sedoglavik's algorithm \cite{sedoglavic2001probabilistic} as noted by \cite{liu2013observability}, in practice some sensors are better than others.}
    \label{fig: cc effective ranks}
\end{figure}

\begin{figure}[h]
    \centering
    \includegraphics[width=\linewidth]{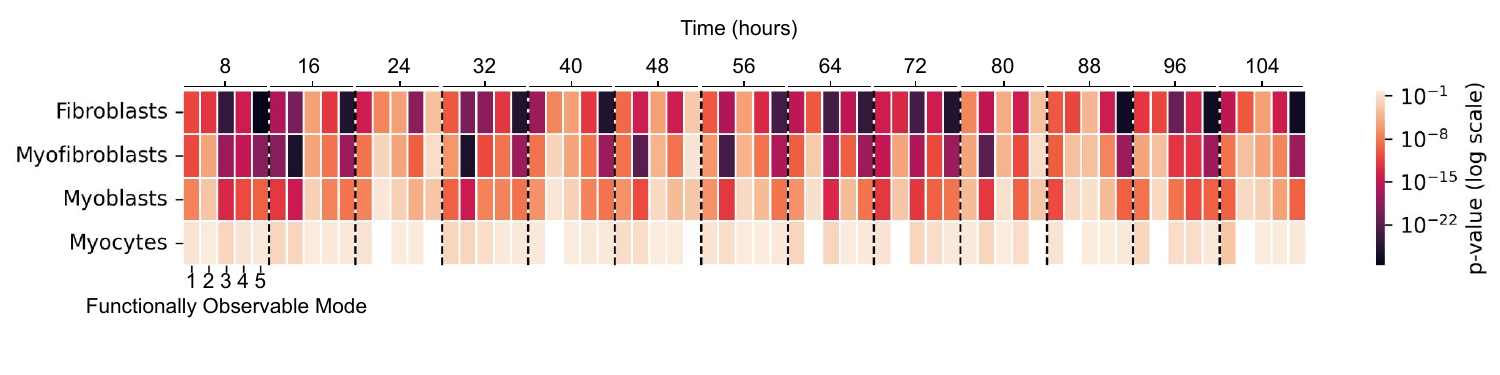}
    \caption{\textbf{Functionally Observable Cell Types.} The top 10\% of genes that contribute to the first 5 functionally observable modes of the observablility matrix $\Os,$ which are obtained as the right singular vectors $\Vv$ from $\Os=\Uv\Sigma\Vv^\top,$ are enriched to identify which cell types are observable \cite{xie2021gene}. In the recreation of Weintraub's reprogramming experiment, Fibroblasts are reprogrammed to myogenic lineages. The functionally observable modes are highly enriched for Fibroblasts, Myofibroblasts and Myoblasts, which are progenitors of Myocytes. Myocytes, which come later in differentiation than Myoblasts, are not functionally observable in this data, which is consistent with the short duration over which the experiment is monitored. The strong enrichment for Fibroblasts and Myogenic lineages indicates that the DSS selected biomarkers make the early stages of reprogramming process functionally observable.}
    \label{fig: cell type enrichment}
\end{figure}

\begin{figure}
    \centering
    \includegraphics[width=\linewidth]{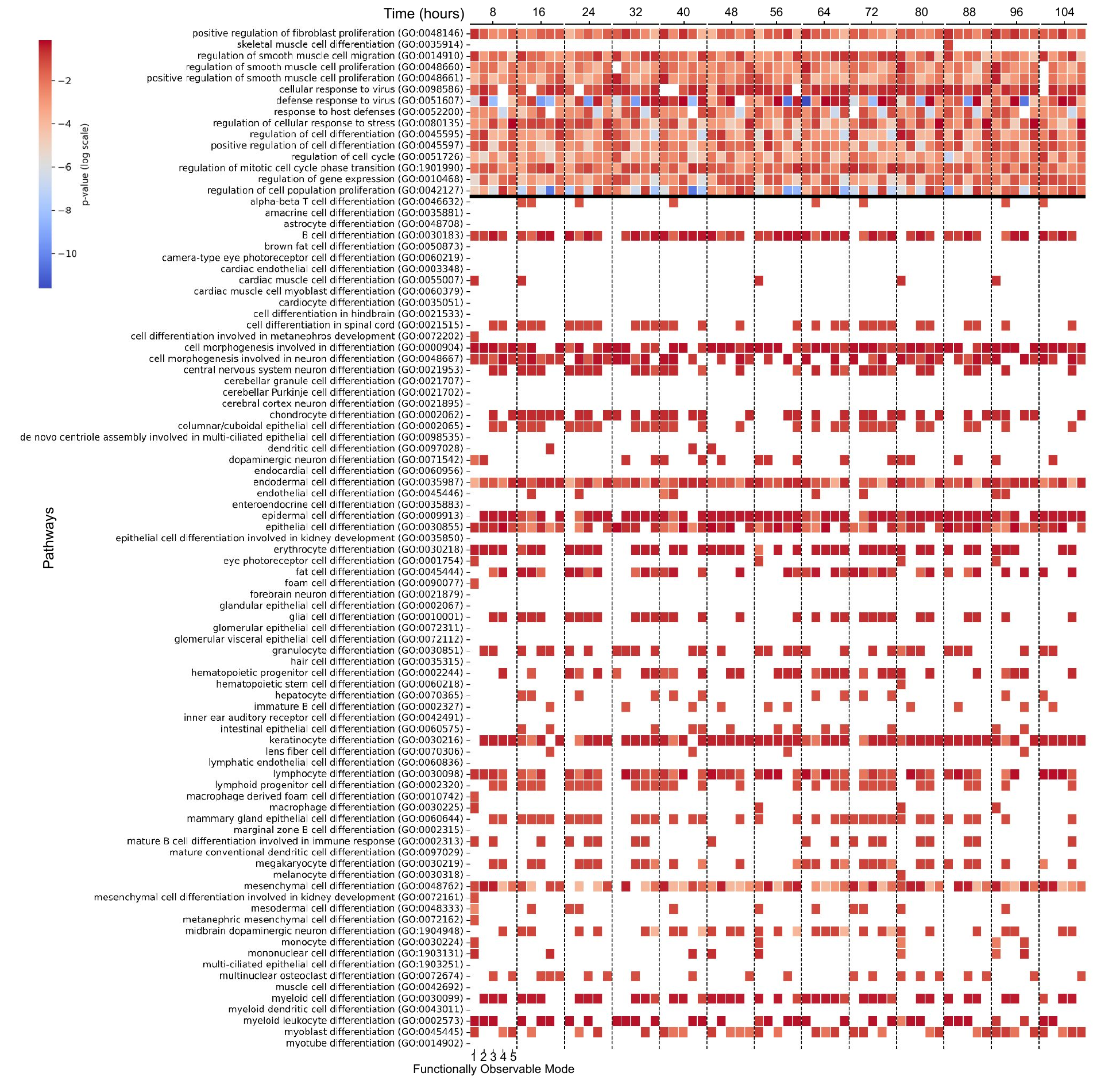}
    \caption{\textbf{Functionally Observable Pathways.} The top 10\% of genes that contribute to the first 5 functionally observable modes of the system are enriched for the related pathways and processes that are monitored by the selected sensors. There is a strong tendency to monitor processes related to myogenic lineage, such as the regulation of smooth muscle cells, and pathways likely involved in the reprogramming process, such as the cellular response to virus, which is expected due to the Lentivirus used to introduce \textit{MYOD} and initiate reprogramming. To contrast the significance of enrichment for myogenic pathways with differentiation of Fibroblasts to other cell types, the enrichment of pathways involved in differentiation to other cell types is shown below the horizontal black line.}
    \label{fig: pathway enrichment}
\end{figure}

\begin{figure}[h]
    \includegraphics[width=\textwidth]{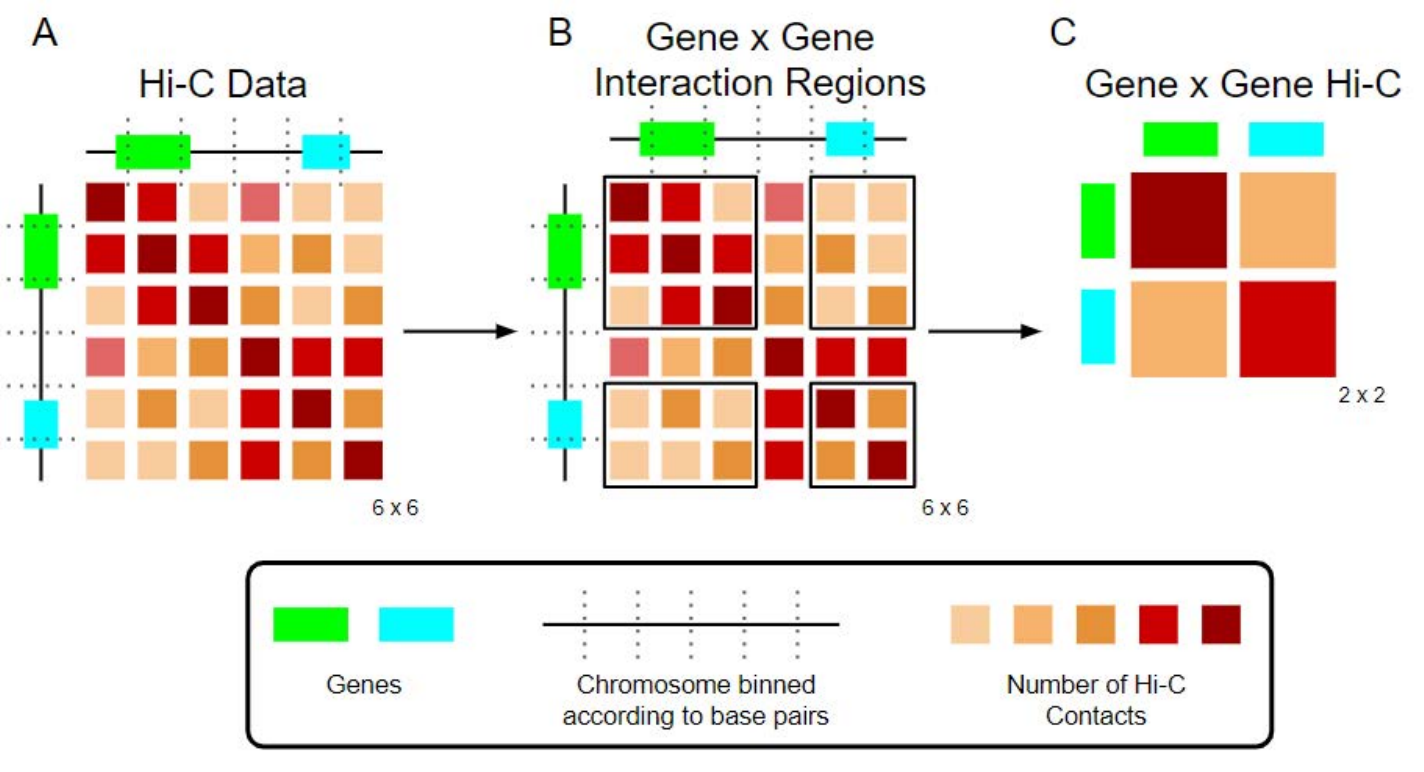}
    \caption{\textbf{Constructing Gene by Gene Hi-C Matrices.} Hi-C matrices, processed to any resolution, is constructed so that each Hi-C index or bin represents a fixed length of the genome (left). Based on the gene coding regions, we can identify the segments of Hi-C corresponding to gene-gene interactions (middle). Averaging over these regions, we can construct gene by gene Hi-C matrices where each row/column corresponds to a single gene and a variable length of the linear genome.}
    \label{fig: geneXgeneHiC}
\end{figure}

\begin{figure}[h]
    \includegraphics[width=\textwidth]{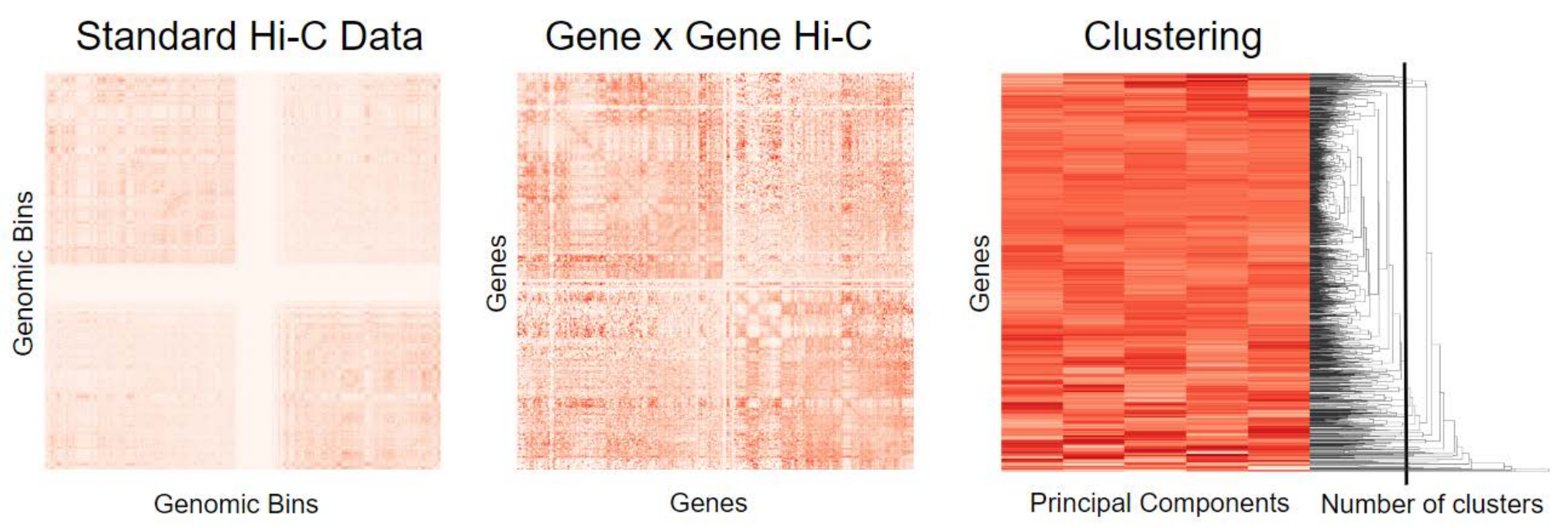}
    \caption{\textbf{Gene Clustering from Hi-C.} Standard Hi-C data from chromosome 1 is shown on the left, where genomic bins correspond to a fixed length of chromatin (i.e. 100kb). The gene by gene Hi-C matrix is constructed according the the process outlined above and in \cref{fig: geneXgeneHiC} (middle). The principal components of the gene by gene Hi-C matrix are used to cluster genes, and the number of clusters is set to maximize the Silhouette score.}
    \label{fig: hicGeneXgene 2}
\end{figure}

\begin{figure}[h]
    \centering
    \includegraphics[width=\linewidth]{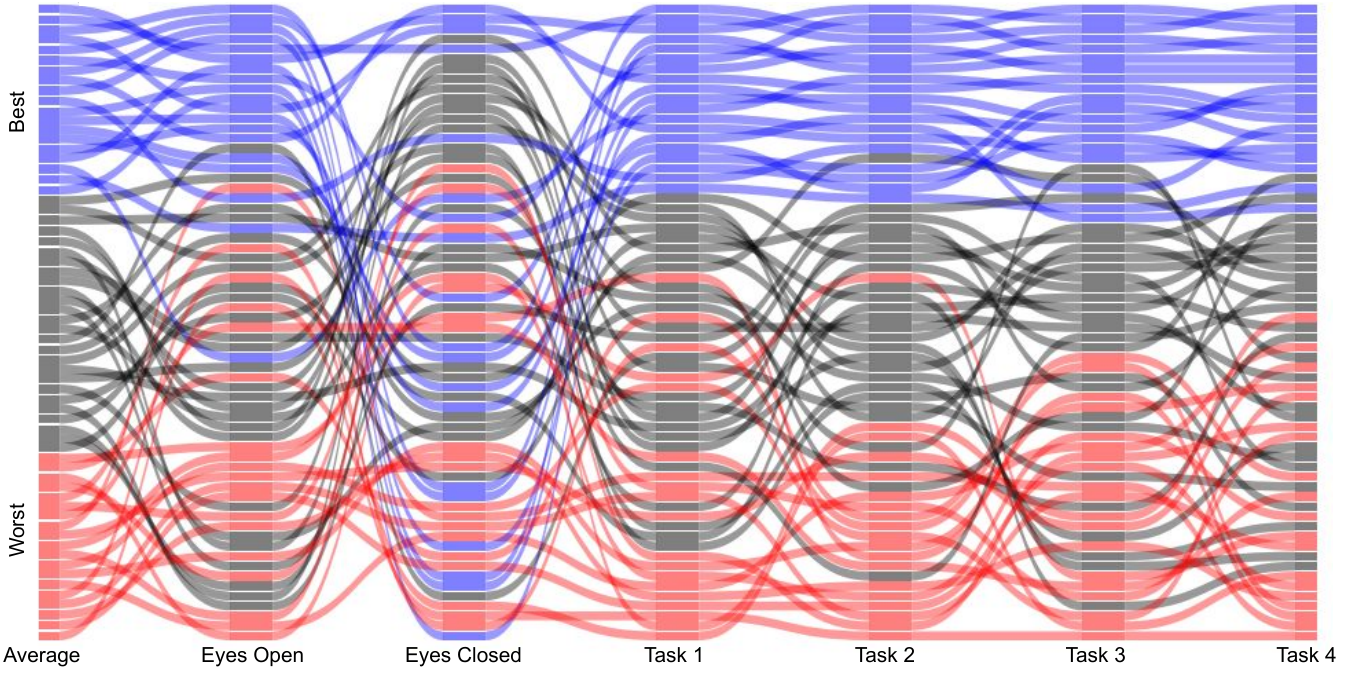}
    \caption{\textbf{Observability Contributions from EEG Signals During Six Tasks.} EEG Signals are ranked according to their contribution to observability. This is an extension of Fig 3.F that shows an additional four tasks. Task 1: open and close left or right fist. Task 2: imagine opening and closing left or right fist. Task 3: open and close both fists or both feet. Task 4: imagine opening and closing both fists or both feet.}
    \label{fig: full EEG data}
\end{figure}

In the strictest sense, the rank of a matrix is defined as the number of its nonzero singular values.
However, due to numerical imprecision and noise in data, it is often more practical to consider only those singular values that are significantly different from zero \cite{konstantinides1988statistical, udell2019big}.
To asses the rank of only significant singular values, several mechanisms thresholding singular values or determinations of effective rank have been proposed \cite{roy2007effective, gavish2014optimal}.
Here, we define the effective rank of a matrix by thresholding its singular values, retaining only those that contribute at least \(\varepsilon\) to the cumulative distribution of singular values.  
The normalized distribution of singular values, \(p_1, \dots, p_k\), is given by  
\[
p_i = \frac{\sigma_i}{\sum_{j=1}^{k} \sigma_j},
\]
where \(\sigma_i\) are the singular values of the matrix.  
The effective rank is then determined as the largest index \(i\) such that \(p_i \geq \varepsilon\):  
\[
\text{Effective rank} = \arg\max_{i} \{ p_i \geq \varepsilon \}.
\]
The distributions of normalized singular values and the effective ranks averaged across all thousand simulated states are shown in \cref{fig: cc effective ranks}.

The effective rank of $\Os_i(\xv)$ depends on both the sensor variable $i$ and the state variable $\xv$.
In \cref{fig: cc effective ranks}, the effective rank averaged over all thousand vectors $\xv$ is shown for each sensors.
Some variables, such as ``mass", are uniformly poor sensors, as shown by the sharp elbow in their distribution of normalized singular values where only $\sigma_1$ has a meaningful contribution.
Other sensors, such as \textit{G2R} are in general very good sensors, with high effective ranks on average (thick black line), but elbows in the distribution show that there exist state vectors $\xv$ for which those sensors make the system less observable.
The variability in the singular values of $\Os_i(\xv)$ indicate that some sensors are better than others for different states $\xv,$ even if the utility of most sensors is theoretically equivalent.

\paragraph{Greedy Sensor Selection.}
The Greedy \cref{alg:greedy} iteratively selects sensors until the system satisfies the criterion $rank(\Os(\xv))=n.$
The algorithm takes as input observability matrices $\Os_i(\xv)$ where the index $i$ indicates the $i$th variable in the system is used as a sensor.
To select the best sensors, the algorithm examines the performance of different combinations of sensors based on the rank of the observability matrices formed by using groups of sensors togther.
The algorithm initializes the desired sensors $\Ds=\emptyset$ as an empty set, and the candidate sensors $\Cs=\{1,\dots,n\}$ (\texttt{Line 2}).
Groups of sensors are evaluated iteratively (loop over \texttt{Lines 4-8}), with the rank based stopping criteria (\texttt{Line 3}).
The utility of adding a new sensor to the desired group $\Ds$ is evaluated by the marginal increase in the rank of $\Os_\Ds(\xv)$ (\texttt{Line 5}), where
\[\Os_\Ds=\begin{bmatrix}
    \Os_{d_1}(\xv)\\
    \dots\\
    \Os_{d_k}(\xv)\\
\end{bmatrix}\text{ where }\Ds=\{{d}_1,\dots,{d}_k\}.\]
After evaluating each candidate sensor, the sensor with the largest marginal increase in the rank of the observability matrix is included in the desired sensor set $\Ds$ (\texttt{Lines 7-8}).

\begin{algorithm}
\caption{Greedy Sensor Selection}
\label{alg:greedy}
\begin{algorithmic}[1]
\State \textbf{Requires:} $\Os_i(\xv)$ for $i=1,\dots,n$
\State{Let $\Cs=\{1,2,\dots,n\}$ and $D=\emptyset$}
\While{$\mbox{rank}(\mathcal{O}_D(\xv))<n$}
    \For{$s\in \Cs\setminus \Ds$}
    \State{Compute $\Delta(s)=\text{rank}(\mathcal{O}_{\Ds\cup \{s\}}(\xv))-\text{rank}(\mathcal{O}_\Ds(\xv))$}
    \EndFor
    \State{Set $s^{*} = \text{argmax}_{s\in \Cs\setminus \Ds}\Delta(s)$}
    \State{Set $\Ds=\Ds\cup \{s^*\}$}
\EndWhile
\State \textbf{Return} The set $D$.
\end{algorithmic}
\end{algorithm}

\newpage
\subsection{Observability of the Andronov-Hopf Oscillator.}\label{SI: empirical gramian}
\paragraph{Andronov-Hopf Oscillator.}
The Andronov-Hopf oscillator is named after the physicist Aleksandr Andronov (1901-1952) and mathematician Eberhard Hopf (1902-1983).
The Hopf bifurcation, where a parameter controls the transition of a stable point to a limit cycle, is also referred to as the Poincar\'{e}-Andronov-Hopf bifurcation to provide attribution to Poincar\'{e} and Andronov for their work studying this bifurcation \cite{poincare1893methodes, marsden2012hopf, marsden1976mathematical, hassard1981theory}.

\paragraph{Empirical Gramians.}
For nonlinear systems, such as the Andronov-Hopf Oscillator, empirical gramians provide a means to asses the observability measures $\Ms_1,\Ms_2,$ and $\Ms_3$ of a system.
While the observability Gramian for LTI and LTV systems can be obtained using the Lyapunov equations or the approximation $\Gv_o\approx\Os^\top\Os$ such approaches are not directly applicable to nonlinear systems.
Instead, the empirical Gramian framework uses simulations to compute observability Gramians \cite{himpe2018emgr}.
The empirical observability Gramian is defined as:
\begin{equation*}
\begin{split}
    \Gv &=\frac{1}{|S_\xv|}\sum_{l=1}^{|S_\xv|}\frac{1}{d_l^2}\int_0\Psi^l(t)\ dt\quad\text{    where  }\quad\Psi_{ij}^l(t)=(\yvt{t}_{li}-\Bar{\yv}_{li})^\top(\yvt{t}_{li}-\Bar{\yv}_{li}).
\end{split}
\end{equation*}
The output trajectories $\yvt{t}_{li}$ correspond go the initial state configuration $\xvt{0}_{li}=d_l\varepsilon_i+\Bar{\xv}.$
See \cite{lall1999empirical, kazma2024observability} for additional details.

In our simulations of the Andronov-Hopf oscillator, we used $\varepsilon=0.01$ and integrated the system for 100 time steps.
We assessed $\Ms_3,$ the trace of the observability Gramian, as a measure of how observable the oscillators were with the fixed parameter.
Since the system is nonlinear, the observability is a function of both the state vector $\xv=\begin{bmatrix}
    x_1&x_2
\end{bmatrix}^\top$ and the parameter $\alpha.$
In general, the observability is more sensitive to $\alpha$ than the initial conditions $\xvt{0}$, but the choice of initial conditions does cause a small amount of variability in the observability of this oscillator, particularly when the system is stable.
%

\subsection{Learning Dynamics from Data}\label{SI: learning dynamics}

\paragraph{Structure of Time Series Data.}
The time series data sets considered in the study were structured in the following form (\cref{tab: data table}):
\begin{equation*}
    \Xv=\begin{bmatrix}
        |&|&&|&|\\
        \xvt{0}&\xvt{1}&\dots&\xvt{T-1}&\xvt{T}\\
        |&|&&|&|\\
    \end{bmatrix},
\end{equation*}
where $n$ denotes the number of state variables (genes, neurons, eeg sensors, etc.) and $T$ is the number of time points.
Several datasets contained multiple replicates, which correspond to multiple time series matrices $\Xv.$

\begin{table}[H]
\footnotesize
    \centering
    \begin{tabular}{lrrrr}
        \toprule
        \textbf{Dataset}&{\textbf{Dimension}}&{\textbf{Time Points}}&{\textbf{Reps.}} &\textbf{Ref.}\\
        \midrule
        \textproc{SBW25} &   624 &  9 & 2&\cite{hasnain2023learning}\\
        \textproc{Reprogramming}&19235& 15 & 3& \cite{liu2018genome}\\
        \textproc{myogenicSignal} &   404 & 15 & 3 &\cite{liu2018genome}\\
        \textproc{MiceNeurons}&21&1508&3&\cite{sweeney2021network}\\
        \textproc{EEG}&64&160&109&\cite{schalk2004bci2000}\\
        \textproc{Proliferation} & 19235 &  8 & 2&\cite{chen2015functional}\\
        \bottomrule\\
    \end{tabular}
    \caption{Time series datasets.}
    \label{tab: data table}
\end{table}

\paragraph{Dynamic Mode Decomposition (DMD).}
DMD finds the best linear operator that explains the data $\Xv.$ It solves the minimization:
\begin{equation*}
\min_{\Av}\|\Xv^+-\Av \Xv^-\|^2_F,
\end{equation*}
where the data $\Xv^-$ and $\Xv^+$ are matrices containing the first and last $T-1$ time points respectively. See \cite{kutz2016dynamic} and references therein for additional details.

\paragraph{Data Guided Control.}
The DGC model for approximating time variant linear systems was proposed by \cite{ronquist2017algorithm} to model the dynamics throughout cell reprogramming. It models the LTV dynamics based on the relationship:
\begin{equation*}
    \Avt{t}=\Iv+\frac{(\xvt{t+1}-\xvt{t})\xvt{t}^\top}{\xvt{t}^\top\xvt{t}}.
\end{equation*}
This modeling approach was proposed to model and control similar dynamics of gene expression. The assumption of the DGC model is that gene expression of a population does not change considerably over time. Hence, the state transition matrix $\Av$ should be similar to the identity $\Iv.$ From this, the authors of \cite{ronquist2017algorithm} define $\Avt{t}$ as a rank one perturbation from the identity to fit the data exactly.

\section{Dynamic Sensor Selection (DSS)}\label{SI: DSS optimizations}
This section outlines further details regarding the observability optimizations of DSS.

\subsection{Output Energy Maximization $\Ms_2$}\label{SI Output Energy Maximization}
Here we provide our method to maximize $\Ms_2$ based on its Lagrangian dual form. We first discuss how this problem is solved when the sensors $\Cv$ are fixed for all time and then consider dynamic sensor selection for $\Ms_2$.

\subsubsection{Time Invariant Sensors}
The objective is to select sensors $\Cv$ that maximize the signal or output energy of the system $\Ms_2$ where
\begin{equation}\label{SI eq Ms2}
    \Ms_2=\sum_{t=t_0}\yvt{t}^\top\yvt{t}=\sum_{t=t_0}\xvt{0}^\top\Gv_0(t,t_0)\xvt{0}.
\end{equation}
Restricting each sensor to measure a single state variable, the optimization is formalized as:
\begin{equation}\label{SI eq cv LTI}
    \max_{\Cv}\Ms_2\text{ subject to }\Cv^\top\Cv=\Iv,
\end{equation}
where $\Iv$ is the identity matrix.
Hasnain \textit{et al.} shows that
the Lagrangian dual formulation of this problem is
\begin{equation*}
    \max_{\Cv}\mathcal{E+L}\text{ where }\mathcal{L}=\textbf{tr}((\Cv\Cv^\top-\Iv)\Dv),
\end{equation*}
where $\Dv$ are the dual variables \cite{hasnain2023learning}. Define the Gram matrix $\Gv$ as
\[    \Gv(t, t_0)= \Phi(t,t_0)\xvt{0}\xvt{0}^\top\Phi(t,t_0)^\top.\]
Following eq. 5 of \cite{hasnain2023learning}, the solution to the optimization problem \cref{SI eq cv LTI} is achieved as:
\begin{equation}
    \dfrac{\partial(\mathcal{E+L})}{\partial\Cv^\top}=2\Gv\Cv^\top-2\Cv^\top\Dv=0,\text{ such that }\Gv\Cv^\top=\Cv^\top\Dv.
\end{equation}
This last expression implies that the eigenvectors of $\Gv$ are the sensor weights or the importance of each sensor at a critical point of the signal output energy with respect to the sensors.
We extend the approach to the selection of time varying sensors.

\subsubsection{Time Variant Sensors}
Selecting time varying sensors $\Cvt{t}$ that maximize $\Ms_2(t)$ at each point in time $t$ can be regarded as solving separate optimization problems at each point in time.
Here, our objective function is
\begin{equation}
    \Ms_2(t)=\yvt{t}^\top\yvt{t}=\xvt{0}^\top\Phi(t,t_0)^\top\Cvt{t}^\top\Cvt{t}\Phi(t,t_0)\xvt{0},
\end{equation}
which is equivalent to \cref{SI eq Ms2} with the removal of the summation over time.
Since $\Cvt{t_{1}}$ and $\Cvt{t_2}$ are independent of one another for all $t_1$ and $t_2,$ we can maximize the energy at each time $\mathcal{E}(t)$ independently of one another.
To do so, we extend the fixed sensor selection optimization of $\Ms_2$ to an optimization of $\Ms_2(t),$ using a similar approach as in \cite{hasnain2023learning}. 
This is formulated as follows:
\begin{equation}\label{SI eq lagrangian dual form for energy}
\begin{split}
    \dfrac{\partial(\Ms_2(t)+\mathcal{L})}{\partial\Cvt{t}^\top} & =\dfrac{\partial}{\partial\Cvt{t}^\top}\bigg(\xvt{0}^\top\Phi(t,t_0)^\top\Cvt{t}^\top\Cvt{t}\Phi(t,t_0)\xvt{0}-\textbf{tr}((\Cvt{t}\Cvt{t}^\top-\Iv)\Dv)\bigg)\\
    & =\dfrac{\partial}{\partial\Cvt{t}^\top}\bigg(\textbf{tr}\big(\xvt{0}^\top\Phi(t,t_0)^\top\Cvt{t}^\top\Cvt{t}\Phi(t,t_0)\xvt{0}\big)-\textbf{tr}((\Cvt{t}\Cvt{t}^\top-\Iv)\Dv)\bigg)\\
    & =\dfrac{\partial}{\partial\Cvt{t}^\top}\bigg(\textbf{tr}\big(\Cvt{t}\Phi(t,t_0)\xvt{0}\xvt{0}^\top\Phi(t,t_0)^\top\Cvt{t}^\top\big)-\textbf{tr}((\Cvt{t}\Cvt{t}^\top-\Iv)\Dv)\bigg)\\
    & =\dfrac{\partial}{\partial\Cvt{t}^\top}\bigg(\textbf{tr}\big(\Cvt{t}\Gv(t,t_0)\Cvt{t}^\top\big)-\textbf{tr}((\Cvt{t}\Cvt{t}^\top-\Iv)\Dv)\bigg)\\
    & = 2\Gv(t,t_0)\Cvt{t}^\top-2\Cvt{t}^\top\Dv=0.
\end{split}
\end{equation}
This has a similar interpretation to the time invariant case and says that the eigenvalues of $\Gv(t,t_0)$ denote the contribution of each state variable to observability at time $t.$
To solve this and select sensors from time $t_0,\dots,t,$ we must form $\Gv(t, t_0)$ for all $t.$ This requires integrating the system forward from the initial conditions $\xvt{0},$ which can be performed efficiently using model reduction, and then computing the largest eigenvector of the matrices $\Gv(t, t_0)$, for which there are fast algorithms.
The weights of the eigenvectors of $\Gv(t,t_0)$ correspond to the contribution to observability provided by each state variable at time $t.$

\subsection{Maximizing the Trace of the Observability Gramian $\Ms_3$}
\label{sec:measures}
Here we provide our method to maximize observability measure $\Ms_3$ with linear programming.
We begin by highlighting a few key properties of the observability Gramian for LTV systems, then provide the integer programming formulation to maximize $\Ms_3.$
Finally, we provide a continuous relaxation of the integer programming problem, which can be solved as a linear optimization.

\subsubsection{Properties of the Observability Gramian}
To formulate the sensor selection problem in terms of integer programming, consider the case where there is a binary variable denoting whether or not each state variable $x_i\in \xv$ is observed or measured at time $t.$ 
Let $\cv_{tj}$ be the $j$-th row of the matrix $\Cvt{t} \in \R^{n\times p(t)}$ such that $\cv_{ij}$ is a row vector with $j$-th entry as $1$ and zero otherwise.
Then the LTV observability Gramian can be written as
\begin{eqnarray*}
  \G_o &=& \sum_{i=0}^{}\Phi(i,0)^\top\Cvt{i}^\top\Cvt{i}\Phi(i,0), \\
   &=&  \sum_{i=0}^{} \sum_{j=1}^{p_k}\Phi(i,0)^\top(\cv_{ij})^\top\cv_{ij}\Phi(i,0).
\end{eqnarray*}
Let $\delta_{tj}\in\{0,1\}$ be a binary variable, which indicates whether $x_j$ variable is measured at time $t$. Then one can express above relation, as
\begin{eqnarray*}
\G_o(\Delta) &=&\sum_{i=0}^{} \sum_{j=1}^{n}\delta_{ij} \Wm_{ij} \text{ where }\Wm_{ij}=\Phi(i,0)^\top(\cv_{ij})^\top\cv_{ij}\Phi(i,0),
\end{eqnarray*}
and, $\Delta=(\delta_{11},\delta_{21},\cdots, \delta_{(T+1)n})^\top\in R^{n(T+1)}$ contains binary variables that indicate which state variables to measure at each time.

\subsubsection{Integer Programming Formulation}
Given the sensor selection variables $\Delta$, the sensor selection problem can be written as a mixed-integer convex problem,
\begin{equation}\label{eq:optint}
    \max_\Delta \Ms_3(t)=\max_\Delta\ trace(\Gv_o(\Delta))\mbox{ subject to} \quad \sum_{j=1}^{n}\delta_{tj}\leq p(t),\quad  
    t=0,1,\cdots \quad \text{where }\quad 0\leq \delta_{kj}\in\{0,1\}.
\end{equation}
Here, the first constraint restricts number of selected sensors to be no more than $p(t)$ for each time point. Because mixed-integer programs do not scale well for large problems, a convex relaxation to (\ref{eq:optint}) provides a useful solution alternative.

\subsubsection{Continuous Relaxation}\label{SI gramian optimization}
In the continuous relaxation, the observation of variable $j$ at time $t$ is relaxed to the interval $\delta_{ij}\in[0,1]$ (i.e. $0\leq\delta_{ij}\leq 1$). This leads to the convex program,
\begin{equation}
\begin{split} \label{eq:optrelax}
    \max_\Delta \Ms_3(t)=\max_\Delta\ trace(\Gv_o(\Delta))\mbox{ subject to} \quad \sum_{j=1}^{n}\delta_{tj}\leq p(t),\quad  
    t=0,1,\cdots \quad \text{where }\quad 0\leq \alphav_{tj}\leq 1.
\end{split}
\end{equation}
The advantage of the relaxation is that it can be solved in time that is polynomial in the number of variables using efficient techniques such as interior point methods. Furthermore, if the solution to the relaxed problem is such that $\alpha_{ij}\in \{0,1\}$ (within numerical tolerance), then the original mixed-integer problem has been solved. The relaxation serves two roles — an approximate (suboptimal) solution to the mixed-integer problem by rounding $\alpha_{ij}$, and, in some cases, a fast optimal solution to the mixed-integer problem.

In both the mixed-integer problem (\ref{eq:optint}) and the convex relaxation (\ref{eq:optrelax}), the desired number of sensors was explicitly constrained to be $p(t)$. Another approach is to allow the number of sensors to be a free variable, and enforce a sparse solution, which can be achieved, for example, by $l_1$ regularization technique which yields a convex problem,
\begin{equation}\label{eq:optrelaxl1} 
    \max_{\Delta} \Ms_3(t) - c\|\Delta\|_1\quad 0\leq\delta_{ij}\leq 1,
\end{equation}
where the constant $c\geq 0$ is the weighting on the $l_1$-norm penalty.
By varying the weight $c$, the number of sensors in the solution set will change to balance the sparsity penalty with the observability measure.

\newpage
\section{Supplementary Figures}\label{SI: figures}

\clearpage

\end{document}